\documentclass[twoside,twocolumn,9pt]{article}
\usepackage{extsizes}
\usepackage[numbers]{natbib}  
\usepackage[left=1.75cm, right=1.75cm, top=2cm, bottom=2.0cm]{geometry}
\usepackage{balance}

\usepackage{sectsty}
\allsectionsfont{\scshape}

\usepackage{graphicx} 
\usepackage{lastpage}
\usepackage{float}
\usepackage{fancyhdr}
\usepackage{fnpos}
\usepackage[english]{babel}
\usepackage{array}
\usepackage[T1]{fontenc}
\usepackage[usenames,dvipsnames]{xcolor}
\usepackage{setspace}
\usepackage{hyperref}
\usepackage[squaren]{SIunits}

\usepackage{upgreek}
\usepackage{xspace}
\usepackage{placeins}
\usepackage{amsmath,amssymb}

\begin{document}

\twocolumn[
  \begin{@twocolumnfalse}
\vspace{3cm}

\begin{center}

    \noindent\Huge{\textbf{\textsc{Experimental models for cohesive granular materials: a review}}} \\
    \vspace{1cm}

    \noindent\large{Ram Sudhir Sharma,\textit{$^{a}$} Alban Sauret\textit{$^{b}$}} \\

    \vspace{5mm}

    \noindent\large{\today} \\

    \vspace{1cm}
    \textbf{\textsc{Abstract}}
    \vspace{2mm}

\end{center}

\noindent\normalsize{Granular materials are involved in most industrial and  environmental processes, as well as many civil engineering applications. Although significant advances have been made in understanding the statics and dynamics of cohesionless grains over the past decades, most granular systems we encounter often display some adhesive forces between grains. The presence of cohesion has effects at distances substantially larger than the closest neighbors and consequently can greatly modify their overall behavior. While considerable progress has been made in understanding and describing cohesive granular systems through idealized numerical simulations, controlled experiments corroborating and expanding the wide range of behavior remain challenging to perform. In recent years, various experimental approaches have been developed to control inter-particle adhesion that now pave the way to further our understanding of cohesive granular flows. This article reviews different approaches for making particles sticky, controlling their relative stickiness, and thereby studying their granular and bulk mechanics. Some recent experimental studies relying on model cohesive grains are synthesized, and opportunities and perspectives in this field are discussed.} \\

 \end{@twocolumnfalse} \vspace{0.6cm}

  ]

\makeatletter
\renewcommand*{\@makefnmark}{}
\footnotetext{\textit{$^{a}$~Department of Mechanical Engineering, University of California, Santa Barbara, California 93106, USA; E-mail: asauret@umd.edu}}
\footnotetext{\textit{$^{b}$~Department of Mechanical Engineering, University of Maryland, College Park, Maryland 20742, USA; E-mail: ramsharma@ucsb.edu}}
\makeatother

\section{Introduction}
\label{sec:sec1_Introduction}

Experimental studies of cohesive granular materials pose a unique challenge. While most fields within mechanics are motivated by a mixture of experimental observations and theoretical explanations, the general simplicity of the interaction of large (\textit{i.e.}, athermal), heavy (\textit{i.e.}, single phase), hard (\textit{i.e.}, participate in perfectly elastic collisions) spheres has allowed substantial progress through particle-resolved numerical simulations and model contact mechanics for cohesionless, and more recently cohesive, particles \cite{guo2015discrete,radjai2017modeling,mandal2020insights,abramian2021cohesion}. Yet, controlled experiments with cohesive granular materials remain more challenging.

Natural contact adhesion between grains can come from various physical sources, such as solid bonds \cite{iveson1996fundamental,mitchell1976fundamentals,goodman1991introduction}, van der Waals forces and electrostatic forces \cite{visser1989van,li2004interparticle}, small capillary bridges formed by the condensation of vapor or the addition of liquid \cite{bocquet1998moisture,halsey1998sandcastles,mitarai2006wet,herminghaus2005dynamics}. Despite mainly limiting such interactions to immediate neighbors, inter-particle cohesive forces can considerably change the force fabric within the granular assembly and, consequently, its bulk mechanics \cite{singh2014effect}. Recent experiments with model contact adhesion between model particles such as glass beads are being used to study granular mechanics from the scale of particles to the scale of bulk processes in a controlled manner. Probing how the magnitude of adhesion, among other particle scale properties, affects bulk behavior of cohesive assemblies remains a major experimental challenge as model approaches for controlled inter-particle adhesive forces are still being developed. In this manuscript, we review the progress in our study of cohesive granular materials through controlled experiments, with a particular emphasis on different model cohesive grains that have been used to conduct experiments in the literature.

\smallskip

\begin{figure*}[h]
  \begin{center}
\includegraphics[width=\linewidth]{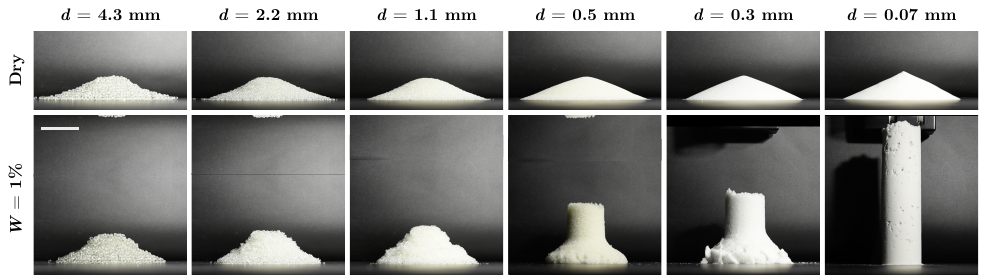}
\end{center}
\caption{Illustration of the effects of grain size and a small amount of water on the macroscopic behavior of granular materials. A mass $m=350$ g of grains is poured into a cylinder, which is then lifted quasi-statically at $0.1$ mm.s$^{-1}$.
Top figures: dry granular columns of the same aspect ratio and different grain sizes collapse into similar piles, described by the macroscopic friction. Bottom figures: When a small amount of water is mixed into the grains, a wide range of behavior is observed due to cohesion.
In this figure, $W= m_{\rm liq}/m_{\rm liq + solid} = 1\, \%$, where $m_{\rm liq}$ and $m_{\rm sol}$ are the masses of the liquid and solid phases respectively for all the grains.
The pictures are taken as soon as the cylinder has been lifted. For large grains, the capillary bridges have only a minor effect. For small grains, the column behaves as a solid and resists any deformation when lifting the confining cylinder. The scale bar shown is the initial diameter of the column, 5 cm.}
\label{fig:Fig1_All}
\end{figure*}

Research to improve our current understanding and modeling of the macroscopic behavior of granular materials comes from the need and desire to control bulk powders and grains for storage and use as described by particle technology \cite{rhodes1990principles,higashitani2019powder}, or through our understanding of soil \cite{verruijt2017introduction} and other natural materials \cite{campbell2006granular,voigtlander2024soft,radjai2017modeling}. In natural granular materials, the distribution of particle properties is often very large, resulting in simplified, statistical descriptions \cite{jaeger1996granular}. The simplifications required to overlay testable properties onto non-idealized granular materials, such as most industrial powders, are often so severe that tested properties are only used to characterize grains. Characterization refers here to the use of empirical tests to differentiate between grain species rather than describe their behavior on the basis of clear physical principles. This last point is precisely the large success of the progress made over the last few decades in describing the static and dynamic properties of model cohesionless granular materials \cite{gdr2004dense}. The discovery of critical parameters in describing the bulk dynamics of granular assemblies resulted in the development of statistical descriptions of bulk flows through some model rheologies \cite{jop2006constitutive,lagree2011granular,kamrin2024advances}. However, results obtained for non-cohesive grains are usually not applicable when a adhesion comes into play, for instance, to account for resistance to deformation due to cohesion. Altogether, this highlights a need for good experimental models of cohesive grains, where the properties of such systems can be studied across length, time, and stress scales. 
\smallskip

Most fundamental studies devoted to cohesive granular materials consider capillary cohesion induced by the presence of liquid owing to its relative experimental simplicity \cite{mitarai2006wet,herminghaus2005dynamics}. The addition of a small amount of water, as shown in Fig \ref{fig:Fig1_All} may be familiar from building sandcastles, can dramatically change the mechanics of a granular material. When wet granular materials are sheared or flow, liquid bridges break and merge, generating localized and heterogeneous effects, making the development of a general framework quite challenging. Reviews specifically on wet grains have been written by \citet{herminghaus2005dynamics} and \citet{mitarai2006wet}, for instance. In the case of powders, which are grains of diameter a few tens of microns and smaller, cohesion may also be a consequence of other forces such as van der Waals or some charging. Delineating the precise cause of adhesion in most cases is not straightforward \cite{royer2009high}, so an important goal is to be able to describe the effects of cohesion at a scale where the physical source of stickiness is not particularly relevant. Altogether, there is a strong need to develop our understanding of the physics of cohesive grains and their collective behavior using a range of physical sources. Being able to experimentally engineer and control the adhesion between particles would pave the way to significant advancements in our understanding of cohesive granular mechanics, such as flow, stability, and erosion. 
\smallskip

Cohesion in powders arising from van der Waals forces is often described from generalizing theoretical derivations, {e.g.}, Derjaguin-Muller-Toporov (DMT) \cite{derjaguin1975effect} or some model experiments developed on Johnson-Kendall-Roberts (JKR) \cite{johnson1971surface}, describing adhesive contact mechanics between particles based on particle stiffness \cite{israelachvili2011textbook}. Such contact descriptions can be used to develop general numerical models for cohesive powders. Such models have also been used to describe how a rheological description of granular flow is modified due to cohesion \cite{mandal2020insights}. The goal of this review is not to cover such models or their consequences but instead to focus on experimental approaches to model and understand cohesion. Such experimental approaches are a step closer to complex assemblies of cohesive particles observed in nature or in industrial processes.

\smallskip

For the sake of consistency, we use \textit{adhesion} to refer to individual stickiness between two grains or between a grain and another surface. We use \textit{cohesion}, similar to fluids, to refer to the general stickiness amongst an assembly of particles. Consequently, for describing individual bonds and sources, we use the phrase \textit{adhesion}, while our primary interest lies in reviewing experiments probing the mechanics of \textit{cohesive} assemblies of grains. Another linguistic difference is often drawn between the definitions of powders and grains. Usually, such a distinction is made based on particle size, with particles smaller than 100 $\mu$m separately called powders. At the scales of powders, cohesion is usually un-ignorable: van der Waals' forces are usually comparable to the particle weight (as shown in Fig \ref{fig:Fig2_Forces}(a)) or as mentioned earlier, humidity and electrical charging can cause such particles to be sticky. For this article, we are interested in the cases where cohesion can be intentionally introduced, so drawing such a distinction between powders and other grains is not particularly helpful here. We do, however, omit discussions of colloids, \textit{i.e.} particles that are small enough to be motivated by thermal fluctuations.
\smallskip

This review is divided into two main parts. In the first part, we review different techniques to induce cohesion in a controlled manner, with attention paid to the advantages and drawbacks of each. In the second part, the observed effects of cohesion in experiments are summarized, both at the particle scale and at macroscopic scales.  A comprehensive review of bulk mechanics with the relevant derivations can be found in most textbooks on the subject, \textit{e.g.}, Refs. \cite{nedderman1992textbook,schulze2008textbook,andreotti2013textbook,hassanpour2019textbook,rumpf1990textbook,duran1999textbook}, with only the relevant characterization discussed here. Following characterization, we briefly summarize the different experimental geometries \& configurations used with model cohesive grains for each scale. Attempts to bridge these two characterizations are discussed. Finally, some perspectives are provided on avenues of future research.
\smallskip

\section{Techniques for Controlling Inter-particle Adhesion}
\label{sec:Control}

\begin{figure}[t!]
\centering
\includegraphics[width =1.05\columnwidth]{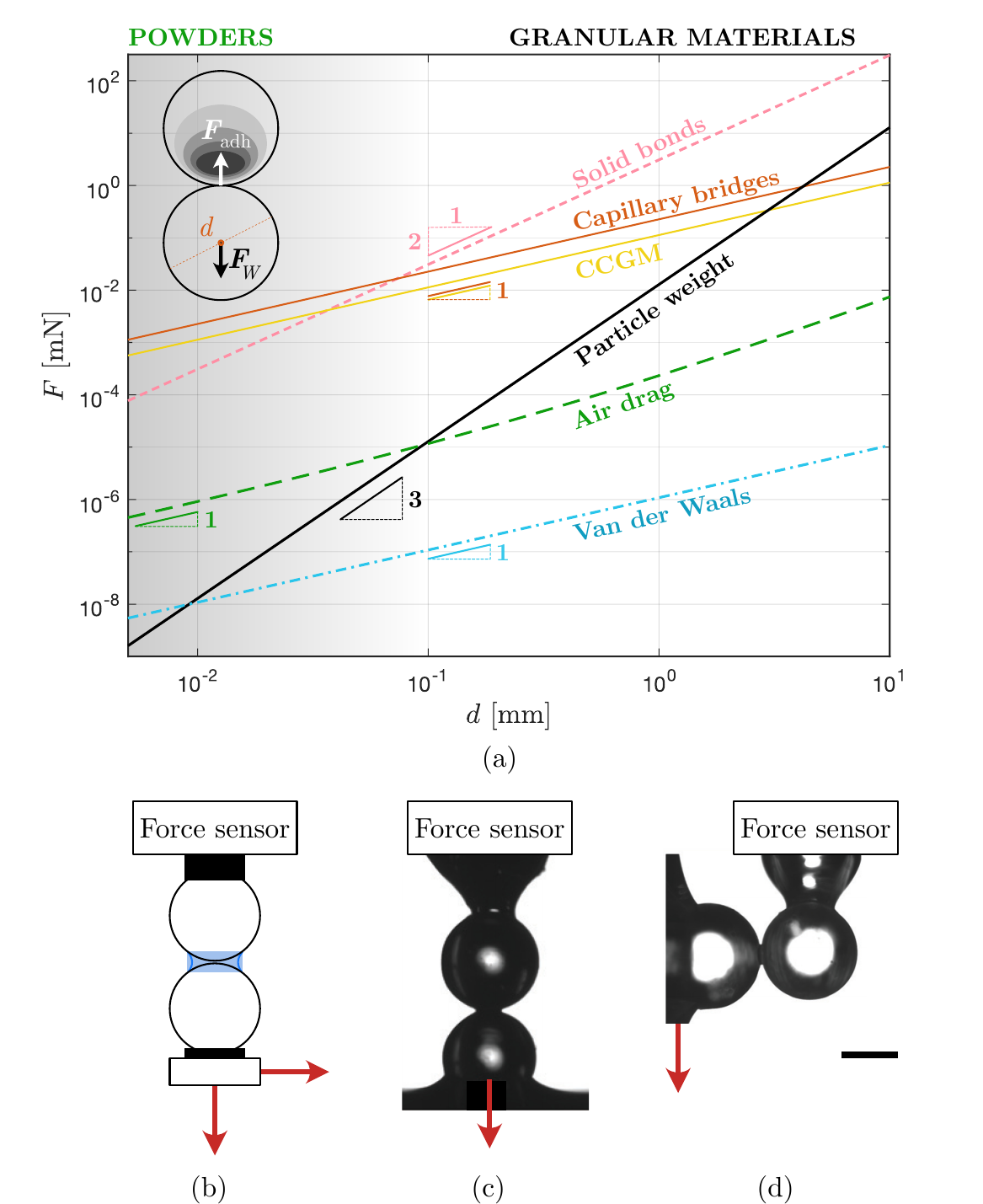}
\caption{(a) Comparison of the order of magnitude of the particle weight, the air drag force, and different interparticle forces when varying the size of a spherical particle of roughness $50\,{\rm nm}$. The expressions used to make this schematic, alongside the magnitudes and descriptions of the constants, are summarized in Table \ref{table:Tab1_Forces} (b) Schematic illustration of micromechanical tests of adhesion between particles in a normal and shear configuration. Pictures of such experiments are shown in (c) and (d), respectively. (c) and (d) are reproduced from \citet{hemmerle2021measuring} with permission from the Royal Society of Chemistry.}
\label{fig:Fig2_Forces}
\end{figure}

The primary goal when making model cohesive materials is to control the force with which neighboring grains are adhering together. Adhesion is a surface phenomenon \cite{israelachvili2011textbook}, and, in the case of powders, for instance, cohesion is a consequence of many surfaces being in close proximity \cite{schulze2008textbook, hassanpour2019textbook, rumpf1990textbook}. Since such forces are not themselves relevant over large distances, they are often described at the scale of individual particles and bonds. As mentioned earlier, several attractive forces can be at play, depending on the size of the particles \cite{brown1970principles,schulze2008textbook}. Figure \ref{fig:Fig2_Forces}(a) reports estimates of the particle weight compared to different inter-particle adhesive forces, and in particular, the models for cohesion that will be discussed in this article. The expressions for the sources of adhesion and the other relevant forces, along with relevant information, are indicated in Table \ref{table:Tab1_Forces}. We also report in figure \ref{fig:Fig2_Forces}(a) an estimate of the air drag experienced by a particle of diameter $d$ in an air flow at $U=1\,{\rm m/s}$: $F_{\rm D}=\pi\,C_{\rm D}\,\,\rho\,d^2\,U^2/8$, where $\rho$ is the density of the air and $C_{\rm D}$ the drag coefficient of a spherical particle \cite{goossens2019review}. The van der Waals force is only significant at very small particle sizes, \textit{i.e.}, for fine powders and colloids. It is directly proportional to the particle size: $F_{\text {vdW }}={A_H d}/({24 h^2})$, where $A_H$ is the Hamaker constant \cite{hamaker1937london}, $d$ is the spherical particle radius, and $h$ is the separation distance between the two particle surfaces, for instance the roughness \cite{israelachvili2011textbook}. When the particles become much larger than $10\, \mu{\rm m}$, the van der Waals forces can be neglected with respect to other cohesive forces and gravity effects. Powders are generally considered as cohesive particles of diameter smaller than $100\,\mu{\rm m}$. Below this diameter, particles are very volatile, and the air drag force is larger than the particle weight. Relying on van der Waals forces to study the role of cohesion is difficult since it requires fine powders where aerodynamic effects are significant, and particle visualization is challenging. Therefore, working with larger cohesive particles, typically larger than a few hundred microns, and model cohesion allows isolating the effect of cohesion from drag force effects. We notice in Fig. \ref{fig:Fig2_Forces}(a) that capillary bridges, solid bonds, and cohesion-controlled granular materials (CCGM) all lead to cohesive force larger than the particle weight and air drag for particles smaller than a few millimeters.

\smallskip

Experimentally, direct measurements are required to measure the adhesive force when separating two grains until the rupture of the bond [Fig. \ref{fig:Fig2_Forces}(b)]. Usually, the maximum tension force measured during the entire test and before splitting is considered as the representative adhesive force $F_{\rm adh}$. Although the most common approach is to measure the tension force when the particles are separated along their axis [Fig. \ref{fig:Fig2_Forces}(c)], other yielding limits such as shear [Fig. \ref{fig:Fig2_Forces}(d)] or torsion can also provide information on the cohesive force and the breaking of the bond \cite{hemmerle2021measuring,farhat2024micro}. The inter-particle adhesive forces lead to macroscopic cohesion, which depends on the amplitude of the force, as well as the number of particles that are in contact with each other, the packing fraction, or the friction coefficient. At the macroscopic scale, one can introduce $\tau_c = \mu \sigma_t$, the cohesive yield stress of the material. For a stress applied larger than $\tau_c$, the cohesive granular material starts deforming plastically. Inside the macroscopic structure, when this happens, the adhesive force between some particles breaks and allows the material to rearrange. The notion of cohesion at the macroscopic scale and the relation with the inter-particle cohesion will be discussed in more detail in section \ref{sec:sec_macro}.

\begin{table*}[t!]
\begin{tabular}{||m{3cm}|m{4.5cm}|m{3cm}|m{0.8cm}|m{2.75cm}|m{1.5cm}||} 
\hline\hline
Sources of adhesion & Expressions for $F_{\rm adh}$ in Fig. \ref{fig:Fig2_Forces}(a) & Model system  & Const. &  Description & Value \\ 
\hline\hline
Capillary bridges     & $F_{\rm adh} = \pi \, \gamma \, d \, {\cos} \theta$  & Liquid, \newline glass spheres \cite{mitarai2006wet} & $\gamma$\newline $\theta$ & Surface tension\newline Wetting angle & $72 \, {\rm mN.m}^{-1}$ \newline $0^{\, \rm o}$\\[0.2ex]
\hline
Solid bridges   & $F_{\rm adh} = k \, \pi \, \sigma_{\rm g, \, p} \, \xi_{p}^{1/2} \, (d/2)^2$  & Solid binder,\newline glass spheres \cite{farhat2024micro} &  $k$\newline $\sigma_{\rm g, \, p}$ \newline $\xi_{\rm p}$ & Numerical prefactor\newline Adhesive strength\newline Relative volume & $1.62$ \newline $275 \, {\rm mN.m}^{-1}$ \newline $0.005$  \\ [0.2ex]
\hline
Polymer coatings   & $F_{\rm adh} = k \, \pi \, \gamma_{\rm PDMS} \, d \, (1 - e^{-b/B})$ & Cross-linked \newline PDMS-OH/H$_3$BO$_3$, \newline glass spheres \cite{gans2020cohesion}  & $k$\newline $\gamma_{\rm PDMS}$ \newline $b$ \newline $B$ & Numerical prefactor\newline Surface tension\newline Coating thickness\newline Roughness related & $1.5$ \newline $24 \, {\rm mN.m}^{-1}$ \newline $1\,  \mu{\rm m}$ \newline $230 \, {\rm nm}$ \\ [0.2ex]
\hline
van der Waals & $F_{\rm adh} = A_{H} \, (d/2) \, / (12 \epsilon^2)$  & Attraction between \newline two spheres \cite{israelachvili2011textbook} & $A_H$ \newline $\epsilon$ & Hamaker constant \newline Surface roughness & $6.5e{-20}$ \newline $50 \, {\rm nm}$\\ [0.2ex]
\hline \hline
Other relevant forces & Expressions for $F$ in Fig. \ref{fig:Fig2_Forces}(a) & Model system  & Const. &  Description & Value \\ 
\hline \hline
Particle weight     & $F_{\rm w} = k \, \rho_g \, g \, \pi \, d^3 \,$ & Glass sphere & $k$\newline $\rho_g$ \newline $g$ & Numerical prefactor\newline Density of glass\newline Gravitational acc. & $1/6$ \newline $2500 \, {\rm kg.m}^{-3}$ \newline $9.8\, {\rm m.s}^{-2}$ \\ [0.2ex]
\hline
Air drag & $F_{\rm drag}=k \, C_d \, \rho_a \, U_a^2 \, A$ \newline\newline $A = \pi \, (d/2)^2$ \newline $C_d=2.7+21.7/\mathrm{Re}+0.1/\mathrm{Re}^2+12.2/\mathrm{Re}^{0.2}-10.6/\mathrm{Re}^{0.1}$ & Drag on a sphere \newline due to air \cite{goossens2019review} & $k$ \newline $C_d$ \newline $\rho_a$ \newline $U_a$ \newline $\mathrm{Re}$ \newline $\nu_a$ & Numerical prefactor\newline Drag coefficient \newline Density of air \newline Velocity of air \newline Reynolds' num. \newline Dyn. viscosity of air & $1/2$ \newline Col 2. \newline $1.2 \, {\rm kg.m}^{-3}$ \newline $0.5\, {\rm m.s}^{-1}$ \newline $U_ad/\nu_a$ \newline $1.6e{-5}\,{\rm Pa.s}$  \\ [0.2ex]
\hline \hline
\end{tabular}
\caption{Expressions of the different sources of adhesion $F_{\rm adh}$ and other relevant forces used in the schematic of forces in Fig. \ref{fig:Fig2_Forces}(a). For each source of adhesion, the relevant constants and the values used as examples to plot the expression in Fig. \ref{fig:Fig2_Forces}(a) are provided. Note other expressions to describe these forces are also possible, these are used here to show the relative magnitudes for different grain sizes.}
\label{table:Tab1_Forces}
\end{table*}

\subsection{Capillary bridges}

The most extensively studied model of cohesive granular material is unsaturated wet grains due to its simplicity in preparation and prevalence in various applications. Adding a small amount of liquid is enough to profoundly change the mechanical properties of a granular material, as shown in Fig. \ref{fig:Fig1_All}. This cohesive behavior results from the formation of capillary bridges between grains, which can be sufficiently strong to form sandcastles \cite{hornbaker1997keeps,pakpour2012construct}. There have been various review papers devoted specifically to wet granular materials, e.g. \cite{mitarai2006wet,herminghaus2005dynamics},, and we here only recall the main results to use them as model cohesive material in the context of this review.
\smallskip

When using wet grains as a model material, one must consider different factors. Various interstitial fluids have been used in experiments, but the choice of the fluid can significantly influence the dynamics of the granular media. The most common fluid used is water due to the simplicity of preparing the material, handling it, and then drying it to retrieve a dry granular material. Glycerol, or mixtures of water and glycerol, is another interstitial fluid frequently utilized in experiments. The mixture allows studying the influence of fluid viscosity on the dynamics of the wet, cohesive granular material. One issue when using water, or mixtures with water, is that one must be careful with evaporation as the fraction of water in the macroscopic granular material will evolve over time. In addition, the evaporation of the liquid can bring some heterogeneities in the material as the evaporation starts from the free surface of the granular material. As a result, silicone oil and mineral oils are also used, in part to avoid evaporation but also due to their controlled rheological properties.

\begin{figure}[t!]
\centering
\includegraphics[width =1.05\columnwidth]{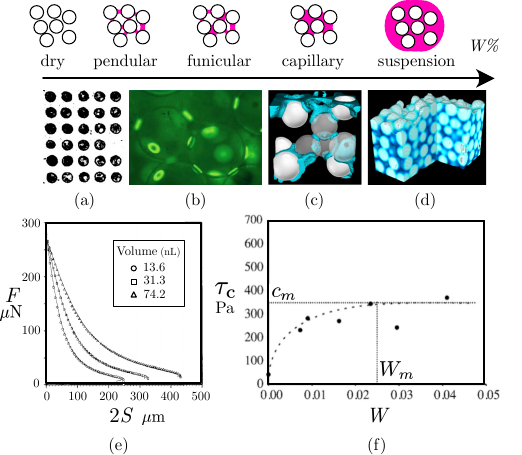}
\caption{Schematic representation of the distribution of liquid within a wet granular material as a function of the liquid fraction $W$ and corresponding examples: (a) dry grains; (b) pendular state [reproduced with permission from \citet{herminghaus2005dynamics}. Copyright 2005 from Taylor \& Francis]; (c) funicular state; (d) capillary state [reproduced from \citet{sauret2020erosion}]. If the liquid fills all the pore spaces and makes a suspension, the effects of cohesion are no longer observed, and this state is consequently outside our interest here. (e) Evolution of the capillary force with the distance between the particles and different volumes of capillary bridges [Adapted with permission from \citet{willett2000capillary}. Copyright 2000 American Chemical Society.]. While more liquid ensures a bridge can last for a greater separation length, the maximum force appears to be approximately the same. (g) Cohesive yield stress $\tau_{\rm c}$ measured with a shear cell when varying the liquid fraction $W$ [Reprinted with permission from  \citet{richefeu2006shear}. Copyright 2006 by the American Physical Society.
]. The dashed line is drawn to suggest that beyond some minimal amount of liquid content, $W_m$, the yield stress is observed to saturate to $c_m$, for the ratio of liquid to total mass tested here.}
\label{fig:Fig3_Capillary}
\end{figure}

To prepare a wet granular material with controlled cohesion, achieving a homogeneously mixed granular material is essential. A possible method involves using a stand mixer, where the grains and the liquid are initially poured with the desired proportion and then mixed thoroughly for a sufficiently long time. The total amount of humidity is often described by the volume or mass ratios of the solid and liquid phases. For Figs. \ref{fig:Fig1_All} and \ref{fig:Fig3_Capillary}, a mass ratio $W= m_{\rm liq}/m_{\rm liq + solid}$, where $m_{\rm liq}$ and $m_{\rm sol}$ are the total masses of the liquid and solid phases mixed respectively. One potential issue when using water or volatile liquids is that they can evaporate and alter the macroscopic cohesion. For situations where evaporation is a concern, grains, and liquid can be mixed within a sealed, rotating drum or a tube placed on a roller mixer. In this case, weighing the container before and after mixing can help verify the amount of liquid adhering to the wall of the container to make sure to start with the desired value of $W$.
\smallskip

At the microscopic scale, the capillary bridge force depends on the physical properties of the particles (size and shape, wettability, roughness, etc.), the fluid properties (surface tension, viscosity, volume, etc) as well as the ambient conditions (pressure, temperature) \cite{halsey1998sandcastles,butt2009normal}. The adhesive force due to a capillary bridge of two spherical grains of radius $R$ in contact and connected by a capillary bridge is:
\begin{equation}
F_{\rm adh}=2 \pi \gamma R \cos \theta,
\label{eq:eq_Adh_cap}
\end{equation}
where $\theta$ is the contact angle of the liquid on the grains. This relation shows that the particle size $R$ plays a critical role in determining the cohesive behavior of granular materials, as the force is directly proportional to the particle radius. Since gravity effects decrease with the volume of the particles, small particles exhibit a much stronger relative cohesion.

Eq. \eqref{eq:eq_Adh_cap} provides the amplitude of the inter-particle cohesion and can be used to estimate the bond number at the particle scale. In a static or flowing granular material, not all capillary bridges have the same volume and are between particles in contact, and some particles move relatively to each other. At the particle scale, different studies have generalized Eq. \eqref{eq:eq_Adh_cap} to the case of spheres separated by a distance of $D$, showing, in this case, a weak dependence on the volume of the capillary bridge, as illustrated in Fig. \ref{fig:Fig3_Capillary}(e) \cite{fisher1981direct,maugis1987adherence,pitois2000liquid}.

The relative motion of the grains leads to the deformation of capillary bridges, which in turn induces viscous forces that resist the movement of particles within it. Different models have been developed at the particle scale to account for the fluid viscosity, particle size, separation velocity, and the dimensionless distance between the grains \cite{ennis1990influence}.

We should also emphasize that, contrary to various numerical models, an important characteristic of the cohesion introduced by liquid bridges is its hysteretic nature. Indeed, a capillary bridge forms between two grains when the liquid films covering them come into contact and coalesce. This occurs at very short distances, on the order of surface roughness. However, the presence of the capillary bridge, and thus some force, is maintained until it breaks, which only happens at distances comparable to the size of the grains, as can be seen in Fig. \ref{fig:Fig3_Capillary}(e). Various parameters govern this distance,  and in particular, the volume of the liquid bridge \cite{maugis1987adherence,simons1994analysis,simons2000direct}. The viscosity of the liquid and the separation velocity can also impact this rupture length \cite{pitois2000liquid,zhao2019capillary}. In addition, for a given volume of the capillary bridge, the contact angle of the liquid wetting the particle can exhibit some pinning during the relative motion of the particles and thus some variation of the capillary force \cite{de2008effect,odunsi2023hysteretic}.

\smallskip

Capillary interactions have been characterized at the scale of two grains for Newtonian fluids, but taking them into account in a granular packing is more challenging. The heterogeneous spatial distribution of the liquid locally modifies the mechanical properties of the wet granular material. The different states of a wet granular material are, in ascending order of liquid fraction, the dry state (without liquid),  pendular state, funicular state, and capillary state, as illustrated in figure \ref{fig:Fig3_Capillary}(a)-(d) \cite{mitarai2006wet,herminghaus2005dynamics}. The pendular regime occurs at low liquid contents, typically below 5\% by mass for monodisperse glass beads. It is characterized by the formation of isolated liquid bridges between particles rather than a continuous liquid phase. The network of liquid bonds induces capillary attraction, leading to a macroscopic cohesion $\tau_{\rm c}$, as shown in Fig. \ref{fig:Fig3_Capillary}(f) \cite{radjai2009bond}. The funicular state occurs at water fractions (per mass) ranging from approximately 5\% to 20\%. Within this range, the liquid content is sufficient to enhance cohesion through the formation of liquid bridges while still allowing for the coexistence of both liquid-filled pores and dry pores within the material. The capillary state for an assembly of glass beads typically lies between approximately 20\% and 40\% of liquid by mass \cite{louati2017effect}. Beyond this, a suspension is obtained, and no more cohesion is observed \cite{hodgson2022granulation}. 

Different studies have considered the macroscopic shear strength properties of wet granular materials in the pendular state. In particular, Richefeu \textit{et al.} \cite{richefeu2006shear} have investigated how water content affects the shear strength of wet granular materials in the pendular state through a combined experimental and numerical approach [Fig. Fig. \ref{fig:Fig3_Capillary}(f)]. They have shown that shear strength primarily depends on the distribution of liquid bonds, resulting in a saturation effect as water content increases and provided an expression for the cohesive stress, similar to the one first proposed by Rumpf:
\begin{equation}
\tau_c=\frac{3 \mu \phi Z F_{\rm adh}}{2 \pi d^2}
\label{eq:eq_Adh_cap_macro}
\end{equation}
where $Z$ is the average number of bonds per particle, $\mu$ the friction coefficient, $\phi$ the solid fraction, $d$ the particle diameter. 
\smallskip

The choice of the liquid and the granular material controls the overall cohesion. However, one could also introduce different levels of cohesion between different populations of beads. For instance, Li and McCarthy have used binary mixtures of spherical particles where some particles are more hydrophobic than others, leading to variations in inter-particle cohesion \cite{li2005phase,jarray2019cohesion}. Using this system allowed them to study the interplay between cohesive forces and particle size distributions in determining the mixing and segregation behavior of cohesive granular materials.

\smallskip

Finally, although an additional phase is present, we should mention the possibility of making immersed granular materials with controlled inter-particle cohesion. Indeed, one can make capillary suspensions, which consist of particles dispersed in a first liquid phase with a small amount of a secondary liquid that is immiscible with the continuous phase. The secondary liquid creates the capillary bridges between the grains, leading to cohesive effects \cite{gogelein2010controlling,bindgen2022behavior,brunier2020generalized}.

\subsection{Solid bridges}

\begin{figure}
\centering
\includegraphics[width =1.025\columnwidth]{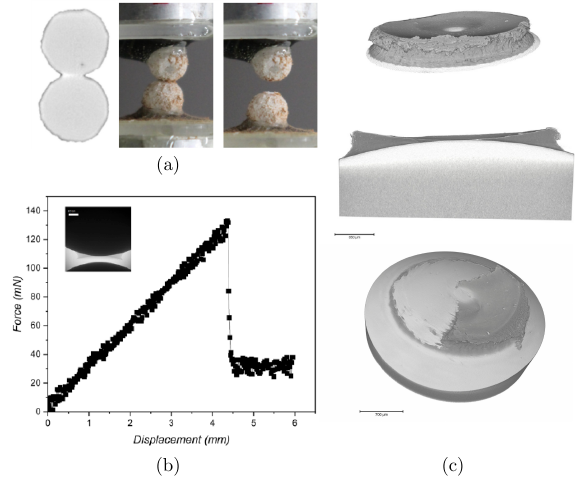}
\caption{(a) Experiments showing microbial-induced calcite precipitation (MICP) to cement grains for tension tests as described earlier [Reprinted with permission from \citet{kwon2024pore}] (b)-(c) Typical set of measurements from particle-scale tensile tests. (b) Evolution of the force between two grains connected by a solid bond. As the distance between rigidly connected spheres increases, the force measured rises until the bond ruptures. The amplitude of the force between before and after the rupture gives the adhesive force. (c) 3D tomographic reconstructions of broken bonds of paraffin wax between 7 mm glass beads. Rupture often occurs as de-bonding from one of the grains (top and middle) or as fracture within the bond (bottom). Figs (b) and (c) are reproduced with permission from \citet{farhat2024micro}, copyright 2024 Springer Nature.}
\label{fig:Fig4_Solid}
\end{figure}

Another experimental method to introduce inter-particle cohesion in a macroscopic granular assembly is through solid bridges between particles, leading to cemented granular materials. In this case, the solid bridges are brittle and do not reform upon breaking. The range of possible cemented granular materials is very large as they can be found in different environmental systems (sandstones, hydrate-bearing sediments, etc.), construction materials (mortars, concrete, and asphalt), sintered ceramics, food industry, etc. Here, we focus on artificial cemented granular materials that have been used in laboratory settings as model cohesive grains.

Even within laboratory-made cemented materials, a wide variety of materials have been used to make solid bonds between glass beads or more angular materials, such as sands. Some of the materials that have been used include salt \cite{soulie2007effect}, paraffin \cite{farhat2024hydraulic}, epoxy \cite{delenne2004mechanical}, polyurethane \cite{ke2025impact}, cement \cite{singh2022evolution}, resin \cite{brunier2020generalized}, or cross-linked PDMS \cite{hemmerle2016cohesive}. Ideally, the amount of cemented materials introduced must remain small, of the order of 1\%-5\% per weight, so that most of the solid bonds are connected to two particles only. One of the main advantages of this approach is that the cohesion does not usually depend on environmental conditions, such as humidity. However, in some cases, the temperature may be a strong parameter.

To prepare a model cohesive granular material with solid bonds, a binding matrix needs to be added to the particles (sand or glass beads). Different approaches can be used to mix the binder and the grains depending on what material is desired to constitute the solid bonds. The most common one is to introduce the liquid binder to the granular material and mix thoroughly to achieve an even distribution. The mixing can be done similarly to the situation with capillary bridges. This mixture of particles and binder is then poured into a container to form the desired shape for further experiments. Then, the liquid binder (cement, clay, PMDS, epoxy resin, etc.) that is initially forming capillary bridging between the particles is left to dry or react such that it hardens after blending. Another approach that can be used is to induce crystallized solid precipitation, often microbial calcite, directly within the granular structure. This matrix can ultimately exhibit varied spatial distributions, from a nearly uniform grain surface coating to highly localized cementing at grain contacts.

Similar to capillary bridges, the characterization of the cohesion force induced by the solid bond can be done between two particles. The tensile force required to separate two spheres connected by a solid bond corresponds to the onset of adhesive rupture [Figs. \ref{fig:Fig4_Solid}(a)-(b)]. Therefore, the amplitude of the force will depend on the properties of the material connecting the particles and the adhesive properties on the surface that depend on the particle (\textit{e.g.}, roughness) and the bonding material. In particular, the rupture of the bond can occur through the detachment of the surface between the solid material and the particle, through the internal failure of the bond, or through a mixed-type bond rupture with some fracture at the bond [Fig. \ref{fig:Fig4_Solid}(c)]. 

Recent experiments have shown that for a paraffin bridge between two millimeter-sized spheres, the tensile force $F_{\rm adh}$ to break the bond can be estimated as
\begin{equation}
F_{\rm adh} =\sigma_{g p} 1.6 \pi \sqrt{\xi_p} R^2
\label{eq:eq_Adh_sol}
\end{equation}
where $\sigma_{g p}$ is the adhesive strength at the interface between the particle and the material making the bond, and $\xi_p$ is the volume content in cemented material. Contrarily to the capillary force, here, the inter-particle force scale with the square of the radius of the particles, which has implications when tuning the size of the granular material.

The cohesion induced by solid bridges is brittle, so the cohesion vanishes upon the breaking of the bond and is not reformed even when particles are in contact again. This approach has been used to study the initial failure of cohesive granular materials. However, similar brittle cohesion has also been used in DEM simulation for a geophysical context to study the collapse of cohesive cliffs \cite{langlois2015collapse}.

Adjusting the size of the grains or the amount of binder put in the initial system can lead to some control over the relative influence of the cohesion. One can also change the nature of the binder to tune the Young’s modulus. Nevertheless, a binding material that does not exhibit a reduction in volume during drying is preferable to avoid pre-stresses and fractures in the cohesive granular material. In addition, depending on the nature of the binder, the solid bridge may not be straightforwardly brittle and may exhibit more exotic mechanics \cite{tardos1996forces,farber2005micro,bika2005strength,kirsch2011measuring}. It is therefore preferable to use paraffin wax or other relatively simple materials to create model binders.

Similar to the previous discussions, different efforts have been made to correlate the microscopic properties of the solid bonds between two particles to the macroscopic mechanical strength. An approach to studying the macroscopic properties of such materials is to perform compression tests and characterize the failure, \textit{e.g.} Refs \cite{zhai2020situ,singh2022evolution,bhat2024micromechanics}. Another one consists of measuring the force required to separate two inverted cones containing the cohesive material. Overall, the results obtained suggest a relationship between the microscopic and the macroscopic solid bond strength, but more investigations on this point remain to be done to obtain a full picture under different loading conditions. From an analytical perspective, the transition from microscopic to macroscopic behavior in a granular material with solid bonds is achieved using homogenization laws, similar to the work developed by Rumpf.

\subsection{Particle cohesion through polymer coating}

\begin{figure}
\centering
\includegraphics[width =\columnwidth]{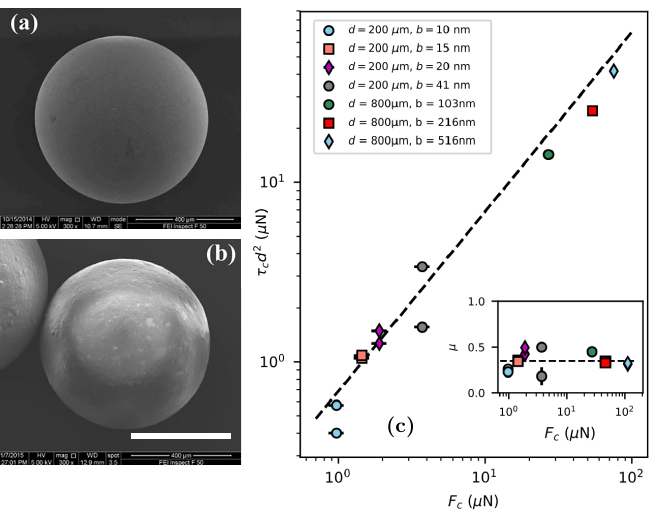}
\caption{SEM images of glass beads (a) before and (b) after polymer coating [Reprinted from 
\citet{ma2019fluidization}, Copyright 2019, with permission from Elsevier].
(c) Macroscopic cohesive stress $\tau_{\rm c}$ of CCGM measured from inclined plane experiments as a function of the interparticle cohesive force $F_{\rm c}$. The dashed line is Eq. \eqref{eq:eq_Adh_cap_macro} with $\phi=0.6$ and $Z=6$, which seemingly provides a reasonable prediction for polymer-coated grains as well. The inset shows that for the range of coatings tested here, no major effect on macroscopic friction $\mu$ is observed. [Reproduced from \citet{gans2020cohesion}, Copyright 2020, with permission from American Physical Society]}
\label{fig:Fig_CCGM}
\end{figure}

A recent approach to producing model cohesive forces between particles relies on a polymer coating of grains of sizes of a few hundred microns to a few millimeters. This approach has been developed by \citet{bouffard2012control}, who demonstrated that the application of polymer coatings leads to an interparticle force, thereby affecting the macroscopic cohesion. The coating consists of an aqueous dispersion of poly(ethyl acrylate) (PEA) and poly(methyl methacrylate) (PMMA) and some binder on granules of cellulose, lactose, NaCl, or sugar beads. The coating is applied by spraying the grains placed in a pheronizer bowl. This polymer coating has then been used by \citet{ma2019fluidization} on glass beads of diameter $d \simeq 600\,\mu{\rm m}$ [Figs. \ref{fig:Fig_CCGM}(a)-(b)]. In both cases, the coating film has a thickness of the order of $5-10\,\mu{\rm m}$. The authors rely on temperature adjustments to tune the interparticle cohesion. As a result, this approach requires the control of both chemical and physical parameters to design a particulate system with desired cohesion characteristics. 

These cohesive granular materials have mainly been used to study the role of interparticle forces on the fluidization characteristics of particles in gas-solid fluidized beds \cite{shabanian2015hydrodynamics,ma2019fluidization} or the flow dynamics in a spheronizer \cite{bouffard2013experimental}. Indeed, one advantage of a polymer coating approach is that once a bond between two particles is broken, it can be restored upon further contact, similar to a capillary bridge. Furthermore, there is no transfer of liquid involved or drainage of the capillary bridge by air flows, which could be detrimental to the long-term evolution of the system.
\smallskip

More recently, another approach relying on a polymer coating has been developed by Gans \textit{et al.}, leading to the development of the so-called Cohesion-controlled granular materials (CCGM) \cite{gans2020cohesion}. The manufacture of CCGM typically involves coating borosilicate glass beads with a polymer that can be adjusted to modify the degree of cohesion. This method allows systematically varying the adhesive forces between particles, providing a versatile platform for experimental studies. The cohesion introduced through this approach remains consistent over time, and the adhesive force can be controlled by adjusting the thickness of the coating. The authors reported no dependence of the cohesion with the temperature in the range $20^{\rm o}C$-$60^{\rm o}C$. This stability is particularly beneficial for experiments that require reproducibility and precision, such as those investigating the flow properties of cohesive materials. Gans \textit{et al.} measured the inter-particle force and reported an empirical expression for the adhesive force between two identical grains \cite{gans2020cohesion}:
\begin{equation}
F_{\rm adh}=\frac{3}{2} \pi \gamma d\left(1-e^{-b / B}\right)
\label{eq:eq_Adh_ccgm}
\end{equation}
where $b$ is the polymer coating thickness, $B \simeq 230 \,{\rm nm}$ for the glass beads considered by Gans \textit{et al.}, is a fitting parameter is of order a few times the particle roughness and $\gamma=24\,{\rm mN.m^{-1}}$, the surface tension of PDMS. 
\smallskip

As discussed in section \ref{sec:sec_macro}, there exists a correspondence between the microscopic inter-particle force and the macroscopic cohesive stress. Gans \textit{et al.} have used an inclined plane setup to investigate the friction and macroscopic cohesion of the CCGM. Their results, reported in Fig. \ref{fig:Fig_CCGM}(c), suggest that the yield stress of the material, $\tau_{\rm c}$ and the microscopic cohesion force $F_{\rm adh}$ are related similar to what was reported for wet granular material $\tau_{\rm c} \sim F_{\rm adh} /d^2$, as illustrated in Fig. \ref{fig:Fig_CCGM}(c) \cite{gans2020cohesion}.
\smallskip

Cohesion-controlled granular materials (CCGM) have been successfully used to study the erosion of a cohesive granular jet by a turbulent flow \cite{sharma2022erosion}, and the collapse of granular columns \cite{gans2023collapse,sharma2024effects}., or the discharge of cohesive grains from a silo \cite{gans2021effect}. Yet, some questions remain to be investigated to fully characterize this model material. In particular, how particle stiffness and tangential forces between particles are affected by the polymer coating remains unclear. Besides, preliminary rheological experiments at varying confining pressures have shown the coating can have effects on the inter-particle friction, in particular for thicker coatings \cite{gans2021rheology}, and this deserves detailed further study.

\subsection{Magnetic forces}

\begin{figure*}
\centering
\includegraphics[width =\textwidth]{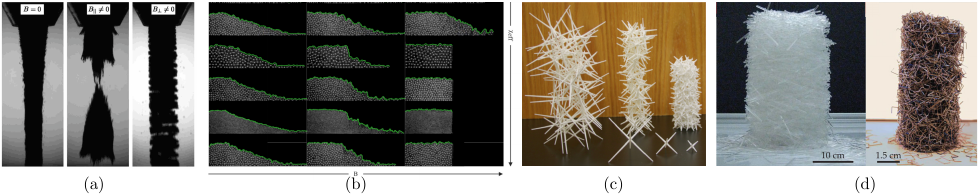}
\caption{Examples of (a)-(b) magnetic and (c)-(d) geometric granular cohesion. (a) Granular jet exiting a funnel in a cohesionless case (left), magnetic field parallel to gravity (middle) and perpendicular to gravity (right) [Reproduced with permission from \citet{lumay2008controlled}, copyright 2008 American Physical Society]. (b) Examples of resulting collapse for different magnetic fields and particles [Adapted with permission from \citet{sunday2024avalanching}. Copyright 2024 American Physical Society]. (c) Columns of 2.5 cm, 5 cm, and 10 cm wide hexapods [Reproduced from \citet{bares2017structure}]. (d) Column of long thin rods (left) and staples (right) maintaining the shape of the container in which they were initially formed [Reproduced from \citet{franklin2012geometric}, with the permission of the American Institute of Physics]}
\label{fig:Fig_Examplesdifferent}
\end{figure*}

A magnetic field applied to an assembly of ferromagnetic beads offers an alternative method to introduce cohesive forces between particles. The magnetic field polarizes the particles, generating poles that interact and produce cohesive/repulsive forces. This approach to cohesion has been considered in different configurations, including the stability of granular slopes \cite{forsyth2001effect,peters2004cohesion,taylor2008influence}, falling granular jet \cite{lumay2008controlled} [Fig. \ref{fig:Fig_Examplesdifferent}(a)], flow in a rotating drum \cite{lumay2010flow}, Janssen effect in a silo \cite{thorens2021magnetic}, or granular collapse \cite{sunday2024avalanching} [Fig. \ref{fig:Fig_Examplesdifferent}(b)], among other.

This method has a few advantages: it is nonintrusive and not very sensitive to environmental conditions. Nevertheless, the granular flow generated using this method differs from typical flow dynamics, as cohesive forces align with the magnetic field, creating directional effects. Additionally, these forces are not evenly distributed around each particle, resulting in anisotropic attractions and repulsions within the material. Moreover, the intensity of the cohesion decreases rapidly with distance from the magnetic source, leading to inhomogeneous and time-varying cohesion as the particles move.

\subsection{Geometrical cohesion}

The different approaches of cohesion presented previously rely on an additional force. However, depending on the particle shape, contact forces and friction can be sufficient to generate geometric cohesion. An example of such cohesive packing consists of randomly packed filaments that can form self-standing structures, similar to a bird nest \cite{weiner2020mechanics} [Fig. \ref{fig:Fig_Examplesdifferent}(c)].

One of the first definition of geometric cohesion was given by Franklin \cite{franklin2012geometric}, following some experiments of jamming \cite{desmond2006jamming} and column collapse of rigid fibers of different aspect ratio \cite{trepanier2010column} [Fig. \ref{fig:Fig_Examplesdifferent}(d)]. The interlocking of fibers can, for instance, prevent granular columns from collapsing, hence demonstrating a notion of cohesion. It also explains why an assembly of flexible fiber presents an elastoplastic response to compression. Interestingly, this notion of cohesion, and thus aggregates, was already observed with fibers by Philipse \cite{philipse1996random} who reported that pouring fibers of very long aspect ratio out of a container produces a plug that retains its shape outside the container thus suggesting an apparent cohesive lengthscale that depends on the granular packing and interlocking.

Although this notion of cohesion is more complex as it involves a crucial role in the shape of the particles, different approaches have been considered to studying particles of concave shape on the static properties of the granular packing. Gravish \textit{et al.} have shown, for instance, that U-shaped particles can form large-scale aggregates by entanglement \cite{gravish2012entangled}. More recently, various studies have considered hexapod-shaped particles and explored how such particles can form column and free-standing structures thanks to the geometric cohesion \cite{zhao2016packings,murphy2016freestanding,zhao2020yielding}, as well as studying the stress and force transmission \cite{athanassiadis2014particle,tran2024contact}.

Few studies have considered flowing properties. Yet, Huet \textit{et al.} have recently considered the avalanche dynamics of cross-shaped particles using a dam break setup and have shown that experimental results are harder to repeat due to the random nature of the entanglement \cite{huet2021granular}. Their study also suggests the presence of mesoscale structures of granular networks of nonspherical particles, which can be seen as reminiscent of a cohesive length.
In a rotating drum, Wang \textit{et al.} have recently shown a regime where concave particles are flowing and one where the particles remain interlocked \cite{wang2024flow}, and laid out a general framework for steady-state behavior \cite{wang2024steady}. They introduced the notion of geometrical consistency to describe how when grains entangle, it causes the bulk material to resist flowing and effectively act as a cohesion.
\bigskip

\section{Effects of Adhesion at the Scale of the Particles}

\label{sec:sec_micro}

\subsection{Characterization}

The weight of the particles is relevant and defined for any particle size [Fig. \ref{fig:Fig2_Forces}(a)]. Therefore, the characterization of the relative effects of the inter-particle adhesion is usually done by comparing the adhesion force to the particle weight. For any source of cohesion, we can define a dimensionless granular Bond number at the grain scale as \cite{nase2001discrete,rognon2008dense,abramian2021cohesion,sharma2024effects}:
\begin{equation}
{\rm Bo}_g = \frac{F_{\rm adh}}{F_{\rm W}},
\end{equation} where $F_{\rm W}$ is the weight of the particle. Note that this definition is the inverse of the regular Bond number used to compare the effects of gravity and surface tension in capillary flows \cite{gennes2004textbook}. Note that some studies use a cohesive number, but the physical meaning remains the same \cite{mandal2020insights,ralaiarisoa2022particle,vowinckel2023investigating}.

While the focus of this article is to review developments of controlled cohesion made through experiments, we should mention that discrete numerical simulations rely on this description. Various numerical models for $F_{\rm adh}$ have been developed and usually rely on a short-range attractive force between particles that rapidly decreases to zero when the grains are separated by an arbitrary distance \cite{rognon2008dense,badetti2018rheology,mandal2020insights,zhu2022grain}. Different reviews and textbooks have summarized some numerical approaches, but cohesive forces used in both Discrete Element Methods (DEM) as well as particle-resolved simulations within fluids often remain ad hoc, containing different assumptions, and are yet to be validated exhaustively \cite{luding2008cohesive,li2011adhesive,10.3390/pr11010005}.
\smallskip

A key advantage of controlled-cohesion materials is the possibility of using large particles, of a few hundred microns to a few millimeters, which allows for more refined imaging than small powders. Furthermore, the large size of particles allows neglecting the effects of air drag, which introduces additional complexities, with respect to the other forces, as can be seen in Fig. \ref{fig:Fig2_Forces}(a) for particles larger than a few hundred microns. To make model cohesion at this scale, the adhesive force must have a magnitude larger or comparable with the weights of the particles. 

Experimentally, the overall effects of cohesion can be increased by either increasing $F_{\rm adh}$ or reducing $F_{\rm W}$. The experimental techniques to control $F_{\rm adh}$ were discussed in Sec. \ref{sec:Control}. Using particles of the same density, tuning the particle size can be used to control the relative effects of cohesion, as shown in Fig. \ref{fig:Fig1_All}, since the weight and the different adhesive forces scale differently with the diameter of the particles. For instance, for capillary cohesion, ${\rm Bo}_g \propto 1/d^2$, decreasing the diameter of the particles by two increases the relative effects of cohesion by a factor of four.

Characterizing adhesion using a granular Bond number is found to be particularly successful when processes are taking place at the scale of individual grains or bonds. These experiments can be categorized into two main kinds. First, where particles are being mechanically detached, and the effect of adhesion is to resist their separation \cite{brunier2020generalized,sharma2022erosion}. Additionally, adhesion modifies the mechanics at the contact affecting regular granular processes such as how particles accumulate into aggregates or piles, and spread.  
\smallskip

\subsection{Detachment and Erosion}

The detachment usually consists of separating two adhesive grains, initially connected. Such experiments are usually used to measure the adhesive force in particle scale tensile tests. Different experimental approaches have been developed to measure such quantities depending on the size of the particles and the adhesive force at play. For glass beads that are large enough, a convenient approach is to use a precision translating stage, where one of the beads is attached, for instance, by epoxy, and one bead is fixed to a support as illustrated, for instance, in Fig. \ref{fig:Fig2_Forces}(b)-(d) and \ref{fig:Fig4_Solid}(a). The force can then be measured by a scale with precision usually of order at least $0.1\,{\rm mN}$ \cite{pitois2000liquid,willett2000capillary,gras2013study,nguyen2021effects,farhat2024micro}. For smaller cohesive force, an alternative is to measure the deflection of a cantilever, which has been calibrated beforehand to obtain the deflection-force curve \cite{he2015measured,lievano2017rupture,hemmerle2021measuring}. Finally, considering very small adhesive forces requires using an Atomic Force Microscopy (AFM) tip to resolve the forces at play \cite{heim1999adhesion,jones2003inter,rabinovich2005capillary,butt2009normal}. Usually, these experiments measure the normal adhesive force through tensile tests, but it is also possible to measure the shear, torsion, bending, and rolling loads \cite{jiang2006bond,farhat2024micro}. Usually, these different quantities exhibit some proportionality between the different yield thresholds.

\smallskip

Additionally, the effects of hydrodynamic forces on assemblies of grains can dislodge particles or, in the case of cohesive grains, erode clusters of particles \cite{besnard2022aeolian,sharma2022erosion}. Experiments have been conducted where hydrodynamic forces are used to break bonds between particles from a granular assembly and cause them to be eroded and suspended. Such experiments provide a simple model for erosion. Experiments with both solid bridges \cite{brunier2020generalized}, or polymer-coated CCGM \cite{sharma2022erosion} have shown that a Bond number can be used to describe modifications on the onset of erosion due to cohesion. Erosion by the impact of other particles, modeled with sand-silicon oil mixtures is to well described by such a description \cite{besnard2022aeolian, selmani2024experimental}. As expected intuitively, more cohesive grains are more resistant to erosion, and the delay in the erosion threshold can be quantified using the Bond number.

\subsection{Effects on Attachment}

Experiments devoted to studying how two grains that are initially separated come into adhesive contact are scarce, as they usually consider multiple grains impacting a cohesive granular bed \cite{pacheco2012sculpting,saingier2017accretion,selmani2024experimental}. A simplified approach for capillary cohesion is to use a Newton's cradle and particles coated with liquids \cite{donahue2008newton,donahue2010stokes2,sekimoto2010newton,donahue2010stokes}. Only recently, a study has considered dry and wet binary collisions of spherical particles using high-speed cameras to capture the results with a collision model to assess dynamics in both dry and wetted conditions \cite{bunke2024three}.

More generally, only a few experimental configurations exist that study the attachment of particles from individuals to clusters of particles of different shapes and sizes. The main experimental approach to study such clustering of particles is either by trapping them at an interface (the "Cheerios" effect) \cite{vella2005cheerios, dalbe2011aggregation, ho2019direct, vassileva2005capillary, huang2012wet} or by acoustically levitating them so as to minimize the effects of gravity \cite{lim2019cluster, brown2024direct, lim2023acoustically}. Experiments with the drop-by-drop accumulation of oil droplets - adhesive but frictionless - into a granular agglomerate or puddle have been shown to display continuum-like behavior \cite{ono2020continuum, hoggarth2023two}.

\section{Effects of Cohesion at Macroscopic Scales}
\label{sec:sec_macro}

Similar adhesive forces between similar particles give rise to cohesion. The techniques of making cohesive grains, as discussed in Sec. \ref{sec:Control}, are designed such that many particles can display similar properties, and their cumulative effects as cohesion can be studied. Indeed, by tuning the cohesion between grains, a wide range of bulk mechanical behavior can be observed, ranging from small cohesion resulting in similar plastic deformation akin to cohesionless grains (as seen in Fig. \ref{fig:Fig1_All} for large grains), all the way to large cohesion resulting in materials that fracture through cracks, akin to amorphous solids (as seen in Fig. \ref{fig:Fig1_All} for small grains). How energy stored in individual bonds contributes to bulk cohesion depends on both the loading history and the given stress state. The local fabric of particles and bonds has a large relevance on any deformation process. Yet, attempts have been made to describe the mechanics of bulk materials, for instance, in the areas of soil \cite{} and powder mechanics, \cite{} where both inter-particle cohesion and broad distributions of particle-scale properties are centrally relevant.
\smallskip

While a distance of about $10$ grain diameters is often used as an order of magnitude used to delineate particle-scale and bulk-scale processes in cohesionless granular materials \cite{andreotti2013textbook}, a similar delineation has not been developed for cohesive grains. In particular, the distances over which the finite-size effects are present in cohesive granular media are expected to depend on the magnitude of cohesion \cite{nowak2005maximum, wu2023combined}. With the assumption that the effects of confinement are negligible, a framework of continuum mechanics is used to describe the material in terms of its relevant stresses \cite{nedderman1992textbook, schulze2008textbook, andreotti2013textbook, hassanpour2019textbook, rumpf1990textbook, duran1999textbook}.
\smallskip

\begin{figure}
\centering
\includegraphics[width =0.85\columnwidth]{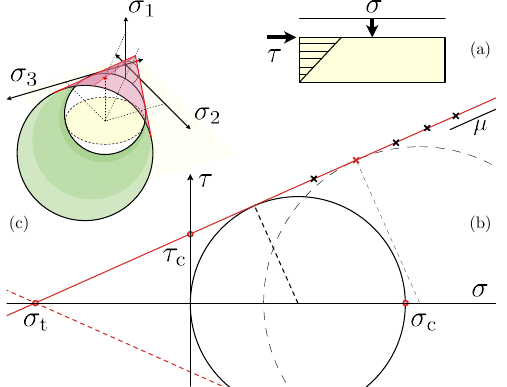}
\caption{(a) A model shear cell, where for a given cohesive granular material, the normal stress $\sigma$ and shear stress $\tau$ are varied until the material yields. (b) Different points of yield (marked `$\times$') are used to estimate the yield curve - displaying the Mohr-Coulomb yield criterion (in red), and measure the stress corresponding to cohesion, $\sigma_{\rm t}$, $\sigma_{\rm c}$ or $\tau_{\rm c}$ as needed. (c) An illustration of the 3D conical yield surface. For the sake of simplicity, the effects of consolidation are not shown here. Colors shown are for illustration purposes.}
\label{fig:Fig3_MohrCoulmob}
\end{figure}

Often, the geometry of the granular assembly is also leveraged to only consider planar stresses, such that they can be represented using a Mohr circle. Derivations of the stresses in this formalism are widely available, for instance, in \citet{nedderman1992textbook}. Combined with a Coulombic-frictional yield criterion, this provides the simplest description of failure and stability in a bulk granular material. For stability, any plane within a cohesionless granular material must have confinement stresses $\sigma$ scaled by $\mu_s$, to be larger than the tangential stresses on the plane $\tau$, $\tau \leq \mu_s \sigma$ \cite{halsey1998sandcastles}. 

The critical angle is related to the internal friction coefficient: $\tan \theta_{c}=\mu_s$. As a column confining a granular assembly is slowly lifted, the confined grains spread into such a sandpile. Experiments are shown in the top row of Fig. \ref{fig:Fig1_All} for a range of grain sizes, initially filled with 350 g of grains, with aspect ratio $a = H_o/R_o \approx 4$. Indeed, in the absence of cohesion, smoothed angles of the relaxed piles are very similar regardless of grain size, $\theta_c \approx 22-23 ^{\rm o}$, i.e. $\mu_s \approx 0.4$.
\smallskip

\subsection{Characterizations of cohesion}

If cohesion is present, the Mohr-Coulomb yield criterion is modified by adding a stress corresponding to cohesion. The stability condition is often written as $\tau\leq\mu_s (\sigma+\sigma_{\rm c})$, where $\sigma_{\rm c}$ is the unconfined yield stress of the material. Alternatively,
\begin{equation}
\tau \leq \mu_s \sigma+\tau_{\rm c}, 
\end{equation}
where $\tau_c = \mu_s \sigma_c$ is sometimes itself called \textit{cohesion} or the yield stress of the material \cite{nedderman1992textbook, schulze2008textbook}. From this formalism, even in the absence of normal stresses, cohesive materials have some intrinsic strength that can resist deformation. The cohesive stresses, $\sigma_c$ or $\tau_c$, are measured for a given cohesion by measuring the normal and shear stresses, $\sigma$, and $\tau$ at the point a granular assembly yields. A schematic illustration of this is shown in Fig. \ref{fig:Fig3_MohrCoulmob}. Such shear cells are common in both academic and industrial research in rectangular \cite{richefeu2006shear,gans2020cohesion,sharma2024effects} and annular geometries \cite{schulze2008textbook} based on the same principles. Measurements at the point of yield for a range of stresses (shown by ``$\times$" in the figure) are used to estimate the yield curve, finding $\tau_c$ (or $\sigma_c$) and friction, $\mu$, shown as a red line. From the analysis, the yield curve is mirrored about the x-axis \cite{nedderman1992textbook}. In 3 dimensions, the Mohr-Coulomb analysis gives rise to a yield surface or locus which is conical in shape \cite{andreotti2013textbook}. Alternate, more nuanced yield criteria can be used and result in modified yield loci, such as the Drucker-Prager or the Cam-Clay models \cite{andreotti2013textbook}. It should be noted that experiments, where the primary stresses can be controlled to confirm the shape of the yield locus, have not yet been performed \cite{andreotti2013textbook}. From the available shear cell tests for the kinds of cohesion discussed above \cite{richefeu2006shear, gans2020cohesion, sharma2024effects},, indeed the coefficient of friction is found to be similar to the cohesionless grains. For large magnitudes of cohesion, such a simple separation of cohesion and friction may not be possible, but additional experiments in this 
region of parameters need to be conducted.
\smallskip

Besides $\sigma_c$ and $\tau_c$, $\sigma_t$ is also used to describe cohesion. All these three quantities are marked for a model yield locus in Fig. \ref{fig:Fig3_MohrCoulmob}(b). The unconfined cohesive stress, $\sigma_c$ is a measure of the systems own compression in the absence of confinement \cite{nedderman1992textbook}. This quantity is often used when compressive stresses are being discussed in the context of cohesive grains \cite{schulze2008textbook, bika2001mechanical}. The yield stress, $\tau_y$, is used to describe the resistance of cohesive grains to shearing stresses \cite{richefeu2006shear}. Finally, tensile stresses in the presence of cohesion are described in terms of the macroscopic tensile strength, $\sigma_t$ \cite{farhat2024micro}. A set of indentation tests, that measure the tensile stress while lifting an indenter, initially compression into a granular bed have also been used to estimate $\sigma_t$ \cite{alarcon2012effect}. In the Mohr-Coulomb model, friction and cohesion are separated from each other, and consequently $\sigma_t \sim \tau_c \sim \sigma_c$. One can readily see how such a description is useful when describing bulk processes. Importantly, such a description is independent of the source of cohesion between grains and is consequently useful to describe powders and soils, where the exact sources of cohesion may be hard to delimit. For the range of grains used in experiments with capillary bridges or CCGM, the cohesive stresses are usually of order a few hundred Pa.
\smallskip

An important purpose of making model cohesive granular materials is to be able to clarify the relationship between the particle- and macroscopic scale characterizations. From the tension-based experiments of particle-particle adhesion and using dimensional analysis, we expect $\sigma_t \sim F_{\rm adh}/ d^2$. Studies have shown that the value of the yield stress $\tau_{\rm c}$ of mono-disperse wet granular material in the pendular regime \cite{richefeu2006shear}, or cohesion-controlled granular material (CCGM) \cite{gans2020cohesion} can also be estimated by relating it to the inter-particle force $F_{\rm c}$, with the same scaling. Rumpf's model enables estimating the macroscopic cohesion using a mean-field method for a homogeneous, isotropic granular medium composed of monodisperse grains while neglecting the effects of the inter-particle force distribution in the material \cite{pietsch1969tensile,pierrat1997tensile,rumpf1990textbook,bika2001mechanical}. The yield stress of the material is then estimated as follows \cite{richefeu2006shear}: 
\begin{equation}
\tau_{\rm c} \simeq \mu\,\sigma_{\rm t}=\frac{3 \,\mu\phi Z \,F_{\rm adh}}{2 \pi d^2}
\label{eq:eq_Rumpf}
\end{equation} 
where $\phi$ is the volume fraction, and $Z$ is the average number of contacts per grain. When grains are cohesive, they can sustain large empty spaces between them and do not pack close to $\phi_{\rm rlp}$ of spheres generally. Additionally, since even experimentally grains are never perfectly monodispersed, a straightforward relationship is not easy to establish between $\phi$ and $Z$ \cite{schulze2008textbook}. As a consequence, despite the appeal of such a relation, it remains to be verified experimentally and carries its largest descriptive power as a scaling relation.
\smallskip

Cohesion is also sometimes characterized by a length scale, which is larger than the scale of the particle and which depends on the amplitude of the attractive forces. From scaling analysis, we expect a length-scale \cite{gans2021effect,gans2023collapse,abramian2021cohesion,sharma2024effects} 
\begin{equation}
\ell_{\rm c} \sim \sigma_{\rm c} / \rho_g \, g,
\end{equation}
where $\rho_g$ is the density of the grain. If compared using $\sigma_{\rm c}$, the corresponding length is an estimate of where compression due to a system's own weight begins to dominate \cite{nedderman1992textbook}. Building a length-scale with $\sigma_{\rm t}$ instead results in an estimate of the length over which tensile stresses can be maintained, which helps explain why such systems readily break up into agglomerates of many sizes and shapes bigger than individual grains. 
\smallskip

It should be emphasized that a flow condition such as Mohr-Coloumb or Cam-Clay is regarding the onset of plastic deformation, and is generally studied in static or quasi-static conditions. Much less is understood systematically about cohesion in dynamic conditions. In the case of cohesion due to capillary bridges, for instance, when inertia is relevant to the deformation of an assembly, other effects, such as the viscosity of the liquid bridge or the lubrication due to the wetting fluid, can modify the apparent cohesion \cite{iveson2002dynamic}. Finally, the effects of cohesion are known to majorly affect how grains pack together, as large void spaces between grains can be left unfilled \cite{fiscina2010compaction,valverde2006random}. Larger cohesion therefore result in smaller overall $\phi$, thereby localizing other mechanical processes.

\begin{figure*}
\centering
\includegraphics[width =0.95\textwidth]{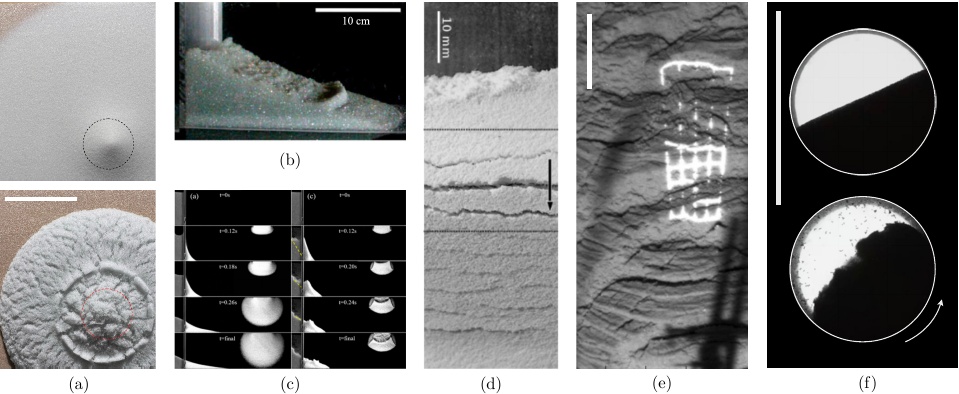}
\caption{(a) The free surface of a pile of grains, after a column of $a = H/R = 5$ is rapidly lifted for cohesionless (top) and cohesive grains (bottom) [Adapted from \citet{sharma2024effects}. Copyright 2024 American Physical Society]. 
(b) The columnar collapse of a cohesive granular column (CCGM) of $a = H/L \simeq 2$ [Adapted from \citet{gans2023collapse}. Copyright 2023 Cambridge University Press]. 
(c) The collapse and unchannelized spread of a column $a = 3$ of dry and wetted grains [Adapted with permission from \citet{li2022unchannelized}. Copyright 2022 American Physical Society]. 
(d) A thin layer of grains, undergoing flexural bending between the two dashed lines, breaks at fixed wavelength [Adapted with permission from \citet{geminard2012flexural}. Copyright 2012 American Physical Society].
(e) Free surface on an inclined granular layer after the flow has stopped [Reprinted with permission from \citet{deboeuf2023cohesion}. Copyright 2023, The Society of Rheology]. 
(f) The free surface of grains in a rotating drum for cohesionless (top) and CCGM for 1 rad.s$^{-1}$. All shown scale bars, unless specified, are 10 cm.}
\label{fig:Fig_ExpCohesions}
\end{figure*}

\subsection{Effects of confinement}

The bulk mechanics, as discussed above, are without any accounting of confining walls. Consequently, the yield criterion defined this way is often called the internal yield locus of the medium \cite{nedderman1992textbook}. To estimate a more realistic yield criterion near confining boundaries, this analysis is often extended to apply a condition of incipient slip at the boundary \cite{hassanpour2019textbook, schulze2008textbook}, for instance, in the design of bunkers \cite{nedderman1992textbook}. Altogether, an \textit{effective yield locus} is often used as a modification at small normal stresses due to the yielding at the walls. However, as suggested earlier, the study of cohesive grains near confining boundaries needs considerable empirical work to explain if slip conditions are indeed similar to the cohesionless case. It is clear that the presence of confinement is an important factor. Often in experiments, then, a layer of grains is stuck onto confining boundaries (fully `rough') to make failure resemble internal yield \cite{nedderman1992textbook,gans2020cohesion,sharma2024effects}.  
\smallskip

Additionally, a \textit{flowability} or flow index \cite{jenike1964report} is often constructed to gauge if a given powder will flow in a given situation \cite{schulze2008textbook,hassanpour2019textbook}. $f$ is described using a comparison of confinement pressure, $\sigma_{\rm applied}$, and the unconfined yield stress, $\sigma_{\rm c}$. Rather than explain the general mechanics of some cohesive grains, such a description is unique to some experimental configurations. In the case that $f \equiv \sigma_{\rm applied}/\sigma_{\rm c} \ll 1$, the material is solid-like, and does not flow. For $f \gg 1$, \textit{i.e.}, $\sigma_{\rm applied} \gg \sigma_{\rm c}$,  the system flows freely as though it were cohesionless. In other words, cohesive granular materials can be made to flow under such conditions.
\smallskip

\subsection{Effects on Free Surfaces}

The most dramatic effects of cohesion are observed in experiments with free surfaces, as shown in Figs. \ref{fig:Fig_ExpCohesions}(a)-(f). The presence of cohesion often acts to preserve the local structure of grains relative to their neighbors, and on a free surface, this results in cracks and gorges many times the size of an individual grain. Perhaps the general fascination relating to sandcastles \cite{hornbaker1997keeps, halsey1998sandcastles, pakpour2012construct, randall2021top} has meant that column stability, angles of cohesive piles and their spreading of collapsing columns have been the most studied set of problems relating to cohesive grains with experiments. 
\smallskip

\citet{pakpour2012construct} provide an expression for the maximum height of a stable column, similar to an expression regarding the buckling of an elastic column, using similarly defined elastic properties for wetted grains. More recently, \citet{gans2023collapse} used CCGM and a length-scale-based description to show the critical height of rectangular columns using the cohesive length $\ell_c$ [Fig. \ref{fig:Fig_ExpCohesions}(b)]. It should be emphasized that the stability of a column is likely highly sensitive to its preparation. The overall packing of cohesive grains is substantially smaller than cohesionless grains, even if some agitation is provided to consolidate after raining in \cite{weis2019structural,wang2021packing}. Consequently, regardless of whether consolidation is provided externally or just due to the weight of the column itself, the average number of contacts may vary \cite{sonzogni2024dynamic}. 
\smallskip

In the same vein, while the maximum angles of stability for cohesive grains are similar in many experimental configurations, some variations are observed for each system. Initial experiments for the stable angle done through the crater left after drainage of wetted grains at the base of a container \cite{albert1997maximum, tegzes1999liquid} allowed for a prediction for the stable angle based on Mohr-Coulomb \cite{halsey1998sandcastles}. The angle of repose has also been studied for a slowly lifting reservoir while measuring the angle of the slowly spreading pile by \citet{samadani2001angle}. \citet{gans2020cohesion} provided a few results on how the angle of repose of CCGM is modified when grains are released from a chute on a round plate. \citet{nowak2005maximum} have provided a framework to estimate the angle in a slowly rotating drum, as it also depends on the drum width. Altogether, larger amplitudes of cohesion lead to larger repose angles, while the precise prediction depends on the pile-making process. While cohesionless grains are smooth to the order of $\sim\, 1d$ for all these measures of the repose angle, cohesion introduces roughness into deposits.
\smallskip

Additionally, experiments with cohesion have been interested in the collapse and spreading of columns of grains, where the confining boundaries are rapidly removed \cite{lube2004axisymmetric, lajeunesse2005granular}. These have been studied using both wetted grains and CCGM for a cylindrical column, spreading axisymmetrically \cite{sharma2024effects} (shown, \textit{e.g.}, in Fig. \ref{fig:Fig_ExpCohesions}(a)); a rectangular column, spreading in a channel \cite{artoni2013collapse, santomaso2018collapse, li2021experimental, gans2023collapse, wu2023combined, sharma2024effects} (shown, \textit{e.g.}, in Fig. \ref{fig:Fig_ExpCohesions}(b)); a rectangular un-channelized geometry \cite{li2022unchannelized} (shown, \textit{e.g.}, in Fig. \ref{fig:Fig_ExpCohesions}(c)). In all cases, more cohesion leads to lesser overall deformation of the column. Some of these studies quantify the magnitude of cohesion for wet grains using the Bond number \cite{artoni2013collapse, li2021experimental,li2022unchannelized}. The concentration of humidity $W_{\%} = m_{\rm liq}/m_{\rm solid}$ plays a large role and is included alongside $\mathrm{Bo}_g$ to explain the range of results as shown by \citet{artoni2013collapse}. Quantifying cohesion using the cohesive yield stress, $\tau_c$, is shown to similarly systematize the spreading behavior of such systems \cite{santomaso2018collapse, gans2023collapse, sharma2024effects}. Indeed, such a description is not specific to the forces at the scale of the grain and has been shown for both CCGM and grains wetted with water cumulatively \cite{sharma2024effects}. 
\smallskip

Cohesive materials usually display roughness on their surfaces. Some examples are shown in Fig. \ref{fig:Fig_ExpCohesions}(a)-(e). A number of studies have looked at the effects of a thin cohesive layer of grains subjected to stresses \cite{alarcon2010softening,alarcon2012effect,geminard2012flexural, tapia2016fracture}. Stretching such layers always leads to instabilities and the formation of patterns on the surface \cite{alarcon2010softening,alarcon2012effect}. In Fig. \ref{fig:Fig_ExpCohesions}(d), we show results from \citet{geminard2012flexural}, where such a thin layer is flexed and undergoes successive fracturing. Further, using the growth of a model crack, \citet{tapia2016fracture} have shown how clusters, dependent on the magnitude of cohesion, control the breakup process. Similar successive structures are observed by \citet{deboeuf2023cohesion} once the flow of a cohesive granular material over a rough inclined plane has finished. Discontinuities are observed in the flowing states as well. Numerical simulations from \citet{abramian2021cohesion} have also shown that roughness exhibited on the free surface after a column has collapsed may be characteristic of the magnitude of cohesion. Finally, grains being rotated in a cylindrical drum, as shown in Fig. \ref{fig:Fig_ExpCohesions}(f) for CCGM, display a wide range of roughness dependent on the cohesion.
\bigskip

Besides their effect on free surfaces, the presence of cohesion is expected to modify other well-studied granular phenomena. The mass flow rate of grains discharged from a bin is reduced by cohesion and is found to depend on the cohesive length in a Beverloo-like relation \cite{gans2021effect}. Similarly, some work on the measurement of drag in unsaturated wet sand \cite{artoni2019drag} and on the impact of projectiles onto beds of wetted grains \cite{zhang2023dynamic} has been performed, but additional studies in different geometries and with alternate cohesions need further consideration.
\bigskip

\section{Conclusions and perspectives}

The different examples described above are a small part of the wide range of granular physics configurations that need to be considered with cohesive grains to further improve our modeling tools. The study of cohesive granular materials has already significantly expanded our understanding of their behavior across various scales. However, even if the past two decades have seen the development of numerous approaches for cohesionless granular flows, cohesive granular systems exhibit even more complex mechanical properties due to inter-particle adhesive forces, which fundamentally alter both micro- and macroscopic behaviors. 

The variety of experimental techniques to make model cohesive granular materials, including capillary bridges, solid bonds, and polymer coatings, now pave the way to exciting development to account for cohesion in granular flow. In this article, we reviewed these different approaches and how they have been used to quantify macroscopic properties such as shear strength, yield stress, and cohesion-related length scales that reflect the effects of interparticle forces. This experimental progress should allow the development of more refined models that bridge particle-scale interactions with bulk material properties, thus offering a framework applicable to both industrial processes and natural phenomena where cohesive particles are significant. In addition, some approaches using cross-linked polymer microparticles are now considered to customize the surface properties in particulate suspensions \cite{moratille2022cross}.

Future research will be able to leverage model cohesive grains to further elucidate different effects of cohesion, such as on the packing density and homogeneity, flow behavior, and responses under various stress conditions. Studies on aging, consolidation, and creep behaviors should be considered, as these time-dependent phenomena introduce complex length-scale dynamics and hysteresis effects that remain underexplored \cite{restagno2002aging,wang2016aging}. In addition, it remains important to establish methods to correlate inter-particle forces with bulk mechanical properties to further enhance modeling approaches for the design and handling of cohesive materials. The dynamics of collisions, breakups, and particle exchange in cohesive systems are still poorly understood, especially for polydisperse and heterogeneous systems where particles have varying degrees of cohesion. For instance, experiments on cohesive particles of varying adhesion could improve our understanding of segregation phenomena \cite{samadani2000segregation}.. Additionally, cohesive rheology remains an exciting avenue for developing continuum models \cite{mandal2020insights,macaulay2021viscosity,gans2023collapse,blatny2024critical}, such as the $\mu(I)$ rheology for cohesionless grains, which relates the shear stress $\tau$ to the pressure $P$ acting on a granular material element through a macroscopic friction coefficient $\mu(I)=\tau/P$, where $I$ is a measure of the system inertia \cite{jop2006constitutive,lagree2011granular,kamrin2024advances}. $I$ also controls the volume fraction, $\phi$. How such rheological descriptions are modified due to inter-particle adhesion has so far only been studied using numerical simulations \cite{mandal2020insights}.  Experimental systems with controlled inter-particle forces also provide valuable benchmarks for numerical simulations, including discrete element methods, enabling comparisons across different cohesion sources and particle types.

\section*{Acknowledgments}
The authors acknowledge James Williams for his valuable input and feedback on the manuscript and the International Fine Particle Research Institute (IFPRI) for their support.

\balance

\begingroup
\small
\setlength{\itemsep}{-1em} 
\bibliography{Biblio} 
\bibliographystyle{unsrtnat} 
\endgroup

\end{document}